\documentclass[12pt]{article}
\usepackage{a4wide,epsfig}

\newlength{\captionwidth}
\renewcommand{\thefootnote}{\fnsymbol{footnote}}
\newcommand{\tMSbar}{\overline{\mbox{\tiny MS}}}
\newcommand{\sMSbar}{\overline{\mbox{\scriptsize MS}}}
\newcommand{\MSbar}{\overline{\mbox{MS}}}

\newcommand{\MOMgg}{\widetilde{\mbox{{\sc mom}}}\mathrm{gg}}
\newcommand{\MOMt}{\widetilde{\mbox{{\sc mom}}}}

\newcommand{\MOM}{\mbox{MOM}}
\newcommand{\MOMggg}{\mbox{MOMggg}}
\newcommand{\MOMg}{\mbox{MOMh}}
\newcommand{\MOMq}{\mbox{MOMq}}

\newcommand{\tMOM}{\mbox{\tiny MOM}}

\newcommand{\sMOM}{\mbox{\scriptsize MOM}}
\newcommand{\sMOMggg}{\mbox{\scriptsize MOMggg}}

\newcommand{\stMOMgg}{\widetilde{\mbox{{\sc mom}}}\mathrm{gg}}

\newcommand{\sMOMg}{\mbox{\scriptsize MOMh}}
\newcommand{\sMOMq}{\mbox{\scriptsize MOMq}}

\catcode`\@=11
\def\slash{\mathpalette\make@slash}
\def\make@slash#1#2{\setbox\z@\hbox{$#1#2$}%
  \hbox to 0pt{\hss$#1/$\hss\kern-\wd0}\box0}
\catcode`\@=12 

\def\bbuildrel#1_#2^#3%
{\mathrel{\mathop{\kern 0pt#1}\limits_{#2}^{#3}}}

\newcommand{\beq}{\begin{equation}}

\newcommand{\eeq}{\end{equation}}

\newcommand{\bea}{\begin{eqnarray}}
\newcommand{\eea}{\end{eqnarray}}

\begin{document}    

\title{\vskip-3cm{\baselineskip14pt
\centerline{\normalsize\hfill Freiburg-THEP 00/12}
\centerline{\normalsize\hfill TTP 00-17}
\centerline{\normalsize\hfill hep-ph/0008094}
\centerline{\normalsize\hfill August 2000}
}
\vskip.4cm
Two Loop QCD Vertices and Three Loop MOM \\
$\beta$ functions 
\vskip.4cm
}
\author{
{K.G. Chetyrkin}\thanks{Permanent address:
Institute for Nuclear Research, Russian Academy of Sciences,
60th October Anniversary Prospect 7a, Moscow 117312, Russia.}
${}^{,a,b}$
and
{T. Seidensticker}${}^{b}$
\\[3em]
${}^a${\it Fakult{\"a}t f{\"u}r Physik,}
\\
{\it Albert-Ludwigs-Universit{\"a}t Freiburg,
D-79104 Freiburg, Germany }
\\
${}^b${\it Institut f\"ur Theoretische Teilchenphysik,} \\
  {\it Universit\"at Karlsruhe, D-76128 Karlsruhe, Germany}
 }
\date{}
\maketitle

\begin{abstract} 
\noindent We present numerical results for the two loop QCD corrections to the
ghost gluon vertex, the quark gluon vertex, and the triple gluon vertex in the 
massless limit at the symmetric point. We restrict ourselves to the tensor 
structures existing in the tree level.  The corrections are used to examine 
different coupling constant definitions in momentum subtraction schemes and 
the corresponding three-loop $\beta$ functions.
\end{abstract}


\thispagestyle{empty}
\setcounter{page}{1}

\renewcommand{\thefootnote}{\arabic{footnote}}
\setcounter{footnote}{0}

\section{Introduction}

There is no unique definition of the QCD coupling constant -- $\alpha_s$: its 
value depends on the renormalization prescription employed. Within pQCD the 
definition which is most often used is based on the $\MSbar$-scheme 
\cite{ms,MSbar}. Such a definition is of great convenience for dealing with
inclusive physical observables dominated by short distances (for a review see
\cite{CKKrep}). On one side the underlying use of dimensional regularization 
makes advanced calculations possible: e.g.~the corresponding 
$\beta^{\sMSbar}$-function is available now with four-loop accuracy 
\cite{beta4l}. On the other one, this makes the $\MSbar$-scheme virtually 
inapplicable in cases when one chooses to utilize a different regularization,
including, most regrettably, the lattice one. In addition, the physical 
meaning of the $\MSbar$ normalization parameter $\mu$ is not transparent and 
leads to the well-known ambiguities when considering the decoupling of heavy 
particles.

These shortcomings are absent for a wide class of so-called momentum 
subtraction ($\MOM$) schemes. The $\MOM$-schemes require the values of 
properly chosen Green functions with predefined $\mu$-dependent 
configurations of external momenta to be fixed (usually to their tree values) 
independently on the considered order. Practical calculations can then be 
performed with any regularization, including the lattice one. 

Unfortunately, even for massless QCD, there are infinitely many possibilities 
to define a momentum subtraction renormalization scheme. Not only is there an 
ambiguity which vertex to subtract, and at which exact configuration of 
external momenta. In addition, there is the freedom to use a certain linear 
combination of the scalar functions appearing in the gluon and quark vertices, 
which can be related to fixed polarization states of the external 
particles \cite{phrva:d24:1369}.

Recently, the momentum subtraction approach has been heavily used to relate 
lattice results for quark masses \cite{NMPmethod,hep-ph/9803491,Becirevic:1999kb} 
and coupling constants 
\cite{hep-lat/9510045,hep-lat/9605033,hep-ph/9810437,hep-ph/9810322,
hep-ph/9903364,hep-ph/9910204,hep-lat/9710044} 
to their perturbatively determined $\MSbar$ counterparts.

In one of these papers~\cite{hep-ph/9903364} it has been argued that the 
knowledge of three-loop coefficients for the corresponding $\beta$-functions 
is necessary and even the four-loop contributions should be taken in to 
account. This is because the accessible energy ranges in these calculations 
are just reaching a level where perturbative QCD calculations start to be 
valid approximations.

The results of these lattice calculations have mainly been analyzed within the 
so-called $\MOMt$-scheme. The scheme is defined using an asymmetric subtraction 
at $q_1^2=q_2^2=-\mu^2, q_3=0$. The perturbative calculations with such a 
definition are certainly {\em much} simpler than those for the schemes 
employing the symmetric subtraction point $q_1^2=q_2^2=q_3^2=-\mu^2$. 
As a consequence  only for that scheme the non-trivial three-loop 
coefficient of the $\beta$-function is known\footnote{Very recently even the 
four-loop term has been computed in \cite{MOMgg:4loop}}. On the other hand, on 
general grounds it is rather clear that $\alpha_s$ defined through the 
symmetric subtraction  point should be less prone to all kinds of 
non-perturbative effects than the one based on the asymmetric subtraction. 

In this paper we suggest a new (approximate) way to compute (massless) vertices
at the symmetric point by using a large momentum expansion. The method reduces 
the problem to evaluation of massless propagators. We compute the two loop QCD
corrections to the ghost gluon vertex, the quark gluon vertex, and the triple 
gluon vertex in the massless limit at the symmetric point in the 
$\MSbar$-scheme. The obtained results are then used to construct different 
coupling constant definitions in momentum subtraction schemes and the 
corresponding three-loop $\beta$ functions.

This letter is organized as follows: In the next Section we discuss the crucial
difference concerning possible non-perturbative corrections to the running
of the coupling constant defined with subtraction at the symmetric momentum 
configuration (which is used in $\MOM$-schemes) and the asymmetric one (used 
in the so-called $\MOMt$-schemes). Section \ref{sec:method} briefly describes
our calculational method. In Section~\ref{sec:vertexcor} the results for the 
vertex corrections at two loops are given. Then we derive the relations between
the coupling constants in the $\MOM$- and the $\MSbar$-scheme 
(Section~\ref{sec:amom}) and close by a short discussion on the effects of our
results on the three loop $\beta$ function in the $\MOM$-scheme 
(Section~\ref{sec:beta}).

\section{MOM vs.~\boldmath{$\widetilde{\mbox{MOM}}$}}

A well-known way to take into account at least some of non-perturbative physics
is the QCD sum rules method \cite{SVZ79} based on the use of Operator Product 
Expansion (OPE) (for a recent review, see \cite{Gubarev:2000if}). For instance, 
let us consider a correlator of two local operators 
\begin{equation}
G^{AB} (q) = i \int \mathrm{e}^{iqx}\mathrm{d} x 
  \langle 0|T\{ A(x) B(0)\}|0 \rangle
\label{AB} 
{}.  
\end{equation} 

At large (and Euclidean) external momentum transfer $q$ the correlator 
(\ref{AB}) can be schematically represented in the form 
\begin{equation}
G^{AB} (q) \bbuildrel{=\!=\!=\!=}_{q \to \infty}^{} G^{AB}_0 (q) 
+ \sum_n C^{AB}_n(q) \langle O_n(0) \rangle 
\label{AB:OPE} 
{},
\end{equation}
where the first term, $G^{AB}_0 (q)$, stands for the purely perturbation theory
contribution while the sum describes the factorization of the non-perturbative 
effects due to the large distances (hidden inside of vacuum expectation values 
(VEV) of various composite operators) and the coefficient functions 
$C^{AB}_n(q)$. The latter correspond to the short distance (of order $1/|q|$)
contributions coming from the integration with respect to $x$ in (\ref{AB})
and, thus, are computable within perturbation theory. 

Next, let us consider a three-point correlator 
\begin{equation} 
G^{ABC} (p,q) = i^2 \int \mathrm{e}^{iqx +i py} \mathrm{d}x \mathrm{d}y
\langle 0|T\{ A(x) B(0) C(y)\}|0 \rangle
\label{ABC:def}
{}
\end{equation}
in the kinematical regime relevant for the MOM-scheme, that is for both $q$
and $p$ being large. In this case one can straitforwardly apply the OPE for 
{\em three} operators and write:
\begin{equation} 
G^{ABC} (p,q) \bbuildrel{=\!=\!=\!=\!=}_{q,p \to \infty}^{} 
G^{ABC}_0 (p,q) + \sum_n C^{ABC}_n(p,q) \langle O_n(0) \rangle 
\label{ABC:OPE} 
{}.
\end{equation}
Here the coefficient functions $C^{ABC}_n(p,q)$ are computable as an expansion
in the (small) coupling constant $\alpha_s(\mu)$ normalized at (large) momentum
scale of order $ |p| \approx |q| $. An important feature of both 
Eqs.~(\ref{AB:OPE}) and (\ref{ABC:OPE}) is that the VEV of composite operators
are universal and do not depend on the correlator. 

The situation is significantly different for the case when only one external 
momentum, say, $q$ is large and another one is fixed or even set to zero (the 
latter case corresponds to a $\MOMt$-scheme). Indeed, in this case the 
representation (\ref{ABC:OPE}) should be replaced with \cite{gOPE}
\begin{eqnarray}
G^{ABC} (p,q) & { \bbuildrel{=\!=\!=\!=\!=}_{q \to \infty}^{}} & 
G^{ABC}_0(p,q) + \sum_i C^{\widetilde{ABC}}_n (p,q) \langle O_n(0) \rangle
\nonumber
\\
&& \mbox{} + \sum_n C_n^{AB}(q) \ i\int \mathrm{e}^{ipy} \mathrm{d}y
\langle 0|T\{ C(y) O_n(0) \}|0 \rangle
{}.
\label{ABC:gOPE} 
\end{eqnarray}
In this representation the first sum comes from the integration region where 
{\em both} $x$ and $y$ are small and of order $1/|q|$. Note  that the functions
$C^{ABC}_n(p,q)$ are in general different from $C^{\widetilde{ABC}}_n(p,q)$.

The second sum results from integration regions of small $ x \approx 1/|q| $ 
only (where the OPE of two operators $A$ and $B$ can be employed). As the 
momentum $p$ is assumed to be small or even zero the two-point correlators in 
the second sum {\em can not} be perturbatively computed and should be considered 
as phenomenological quantities analogous to "condensates".  

Thus, we conclude that the three-point correlators employed in $\MOMt$-schemes
contain, in in addition to the VEV of composite operators, an extra source of
the non-perturbative corrections --- VEV of bilocal operators --- in comparison
to the same correlators with symmetrical pattern of the external momenta.

\section{\label{sec:method}Method of the calculation}

The analytic computation of arbitrary two loop three point functions seems to 
be excluded at the time being. Therefore our computation is based on the method
of asymptotic expansions \cite{Smi95} which reduces the complexity of the 
integrals. We consider three point graphs with incoming momenta $q_1$,
$q_2$, and $-(q_1+q_2)$ and identify one momentum as large, say $q_1$. The 
ordinary large momentum procedure is used to find an expansion with respect to 
$q_2^2/q_1^2$ and $q_1.q_2/q_1^2$. The symmetric point further demands that 
$(q_1+q_2)^2 = q_1^2$, so $q_1.q_2$ can be reexpressed by $q_2^2$. Setting 
$q_2^2=z q_1^2$ yields a series in the variable $z$. 

Using the described method we found very good numerical agreement with the 
analytical results for the one loop vertex corrections given in \cite{CelGon79,davyd96} 
and the one and two loop scalar three point integrals given in \cite{DavUss94}
by computing only 4 to 6 terms of the expansion.

The use of the large momentum procedure results in an asymmetric approach and
marks one momentum. Therefore one obtains at least two independent series 
by choosing different momentum distributions. The numerical values of the 
series must coincide at the symmetrical point ($z=1$). This provides a powerful 
check on our method and can be tested by examining the non-symmetric scalar 
ladder type topology computed in \cite{DavUss94}.

We use the series with the best convergence to estimate the mean value and
the variation of independent expansions as an indication for the size of
the uncertainty. This description is in most cases equivalent to an error
estimation based on the size of the highest expansion term available.

Throughout the whole computation we made use of several computer programs.
The diagrams were generated using {\tt QGRAF} \cite{Nog93} and the large
momentum procedure was applied by {\tt EXP} \cite{Sei:dipl}. The actual
evaluation of the integrals was done using the {\tt MINCER} \cite{MINCER}
package written in {\tt FORM} \cite{FORM}.
 
\section{\label{sec:vertexcor}Two loop vertex corrections}

We parametrize the QCD vertices at the symmetrical point 
($p_1^2=p_2^2=p_3^2=p^2$) as follows:
\begin{itemize}
\item ghost gluon vertex ($p_1$ is the incoming momentum of the out-ghost)
\begin{equation}
  \tilde{\Gamma}^{abc}_{\mu}(p_1,p_2,p_3) = 
     i g_s f^{abc} p_1^\nu 
     \left( g_{\mu\nu} \tilde{\Gamma}(p^2) + \cdots \right),
\label{eq:defccg}
\end{equation}
\item quark gluon vertex
\begin{equation}
 {\Lambda}_{\mu,ij}^a(p_1,p_2,p_3) =  g_s T^a_{ij} 
  \left( \gamma_\mu \Lambda(p^2) + \cdots \right),
\label{eq:defqqg}
\end{equation}
\item triple gluon vertex
\begin{eqnarray}
  \lefteqn{ \Gamma^{abc}_{\mu \nu \lambda}(p_1,p_2,p_3) = - i g_s f^{abc}
\times} \nonumber\\&&\mbox{}
    \left[ 
       \left( g_{\mu \nu} \left( p_1 - p_2 \right)_\lambda 
            + g_{\nu \lambda} \left( p_2 - p_3 \right)_\mu
            + g_{\lambda \mu} \left( p_3 - p_1 \right)_\nu \right)
       \Gamma(p^2) + \cdots 
   \right]
\label{eq:defggg}
\end{eqnarray}
\end{itemize}
with the strong coupling constant $g_s$, the structure constants of the SU(N) 
Lie algebra $f^{abc}$, and the matrices of the fundamental representation 
$T^a_{ij}$. The terms contained in the ellipsis are defined in such a way that
they do not interfere with the displayed form factor. This means that e.g.~for
Eq.~(\ref{eq:defqqg}) the dots could  contain only the structures 
$p_i^\mu, \slash{p_i}p_j^\mu$ and so on. Note, that our choice of the 
kinematical structures (\ref{eq:defccg}-\ref{eq:defggg}) is, probably, the 
simplest one. However, our approach could be applied, if necessary, to any other 
choice, should the latter be more convenient and/or natural for a specific 
physical problem. 

\begin{table}[!tb]        
\begin{center}
\begin{tabular}[t]{|c|c|c|c|}
\hline
& $C_F$ & $C_A$ & $T{} n_f$  \\
\hline
$\tilde{\Gamma}^{(1)}(-\mu^2)$ & --- & $\frac{3}{32} + \frac{1}{96} I$ & --- \\
\hline
$\Lambda^{(1)}(-\mu^2)$   & $- \frac{1}{2} + \frac{1}{12} I$ 
                               & $\frac{13}{16} - \frac{13}{96} I$ & --- \\
\hline
$\Gamma^{(1)}(-\mu^2)$         & --- & $-\frac{3}{32} + \frac{23}{288} I$ 
                               & $\frac{1}{2} - \frac{2}{9} I$ \\
\hline
\end{tabular}
\end{center}
\caption{\label{tab:oneloopindcol}One loop results --- by SU(N) color factors.}
\end{table}

\begin{table}[!tb]
\begin{center}
\begin{tabular}[t]{|c|c|c|c|c|c|}
\hline
& $C_F^2$ & $C_F C_A$ & $C_A^2$ & $C_A T{} n_f$ & $C_F T{} n_f$ \\
\hline
$\tilde{\Gamma}^{(2)}(-\mu^2)$ & --- & --- & $0.197(1)$ & $-0.151(2)$ & --- \\
\hline
$\Lambda^{(2)}(-\mu^2)$   & $0.206(4)$ & $-0.20(4)$ & $0.679(1)$ 
                               & $-0.4968(4)$ & $-0.0211(4)$ \\
\hline
$\Gamma^{(2)}(-\mu^2)$         & --- & --- & $-0.22(4)$ & $0.65(7)$ 
                               & $-0.408(10)$ \\
\hline 
\end{tabular}
\end{center}
\caption{\label{tab:twoloopindcol}Two loop results --- by SU(N) color factors.}
\end{table}

With Eqs.~(\ref{eq:defqqg}) and (\ref{eq:defggg}) we follow the prescription 
of \cite{CelGon79}. Concerning the ghost gluon vertex our definition partly 
avoids the ambiguity connected with the identification of the vertex form 
factors and differs from the one used in \cite{CelGon79}. 

Each form factor receives quantum corrections:
\begin{equation}
  \Gamma = 1 + \frac{\alpha_s}{\pi} \Gamma^{(1)}
             + \left( \frac{\alpha_s}{\pi} \right)^2 \Gamma^{(2)} + \cdots
\end{equation} 
and similarly for $\tilde{\Gamma}$ and $\Lambda$.

In the following we restrict ourselves to Landau gauge so the gluon propagator
is taken to be purely transversal. A further simplification can be achieved
by observing that the $\MOM$-scheme is usually defined at $p^2 = - \mu^2$
with the 't Hooft mass $\mu$. Therefore all logarithms $\log(-p^2/\mu^2)$ 
drop out. All quantities displayed in the following are computed using the
$\MSbar$-scheme for the renormalization.

The one loop corrections to the ghost gluon vertex, the quark gluon vertex,
and the triple gluon vertex can be found in \cite{CelGon79}. There the 
integral 
\begin{equation}
I = -2 \int_0^1 \frac{\log(x)}{x^2-x+1} = 2.3439072\ldots
\label{eq:defI}
\end{equation}
was defined. The results are displayed in table \ref{tab:oneloopindcol}
where we have distinguished between contributions proportional to different 
color factors of the SU(N) (for SU(3) $C_F = 4/3$, $C_A = 3$, and
$T{}=1/2$, $n_f$ represents the number of fermions).  

Our numerical results for the two loop contributions are summarized in table 
\ref{tab:twoloopindcol}. The number in the brackets denotes our error
estimation on the last digit.

Inserting the values of the color factors for SU(3) and linearly adding the 
errors one would surely overestimate the uncertainty. Therefore we first 
computed the series with color factors already inserted and then extracted 
the results which read
\begin{eqnarray}
  \tilde{\Gamma}^{(2)}(-\mu^2) &=& 1.770(9) - 0.1727(3) \, n_f, \nonumber\\
  \Lambda^{(2)}(-\mu^2)        &=& 5.590(6) - 0.7594(2) \, n_f, \nonumber\\
  \Gamma^{(2)}(-\mu^2)         &=& -2.0(4) + 0.72(9) \, n_f.
\end{eqnarray}

\section{\label{sec:amom}Relations between coupling constants}

The $\MOM$ renormalization condition implies that the value of one of the above
defined functions should be equal to 1 when the corresponding $\MOM$-scheme is 
used. This allows for the computation of the coupling constant in that 
$\MOM$-scheme ($\alpha_s^{\sMOM}$) as a perturbative series in the coupling 
constant of the $\MSbar$-scheme ($\alpha_s^{\sMSbar}$) which is of the form
\begin{eqnarray}
  \frac{\alpha_s^{\sMOM}}{\alpha_s^{\sMSbar}} &=&
\left\{ 1 
  + \frac{\alpha_s^{\sMSbar}}{\pi} 
      \left( d_{10} + d_{11} n_f \right)
  + \left( \frac{\alpha_s^{\sMSbar}}{\pi} \right)^2
       \left( d_{20} + d_{21} n_f + d_{22} n_f^2 \right)
  + \cdots
\right\}.
\end{eqnarray}
We will refer to $\MOM$ quantites computed on the basis of the triple gluon
vertex, the quark gluon vertex, and the ghost gluon vertex as $\MOMggg$,
$\MOMq$, and $\MOMg$, respectively.

Using the well known results for the self energies of the quark, the gluon,
and the ghost at one and two loops in the $\MSbar$-scheme we find the 
coefficients shown in table \ref{tab:amom}. 

\begin{table}[!tb]
\begin{center}
\begin{tabular}[t]{|c|c|c|c|c|c|}
\hline
            & $d_{10}$ & $d_{11}$ & $d_{20}$ & $d_{21}$ & $d_{22}$ \\
\hline
$\MOMg$     & $\frac{49}{12}+\frac{1}{16} I$ & $- \frac{5}{18}$
            & $35.88(2)$  & $- 5.0707(6)$ & $\frac{25}{324}$ \\
\hline
$\MOMq$     & $\frac{89}{16} - \frac{85}{144} I$
            & $- \frac{5}{18}$ & $29.53(1)$
            & $- 5.1961(4)$ & $\frac{25}{324}$ \\
\hline
$\MOMggg$   & $\frac{11}{2}+\frac{23}{48} I$
            & $- \frac{1}{3} - \frac{2}{9} I$ & $59.8(8)$
            & $- 12.6(2)$
            & $\frac{47}{432} + \frac{7}{54} I + \frac{1}{81} I^2$ \\
\hline
\end{tabular}
\end{center}
\caption{\label{tab:amom}Coefficients of the relation between the coupling
constants in the $\MOM$- and the $\MSbar$-scheme.}
\end{table}

\section{\label{sec:beta}Beta function}

We define the perturbative series of a generic $\beta$ function by
\begin{eqnarray}
  \mu^2 \frac{\mbox{d}}{\mbox{d} \mu^2} \frac{\alpha_s(\mu^2)}{\pi} 
  = \beta(\alpha_s) = 
  - \left( \frac{\alpha_s}{\pi} \right)^2
  \sum_{i\ge0} \beta_i \left( \frac{\alpha_s}{\pi} \right)^i.
\end{eqnarray}

The $\beta$ function in the $\MOM$-scheme ($\beta^{\sMOM}$) can be computed 
directly from the results presented in the last section using the relation
\begin{equation}
  \beta^{\sMOM}(\alpha_s^{\sMOM}) 
  = \mu^2 \frac{\mbox{d}}{\mbox{d} \mu^2} \frac{\alpha_s^{\sMOM}(\mu^2)}{\pi}
  = \left.
     \beta^{\sMSbar} 
    \frac{\mbox{d} \alpha_s^{\sMOM}}{\mbox{d} \alpha_s^{\sMSbar}}
    \right|_{ \alpha_s^{\tMSbar} \to \alpha_s^{\tMOM}}.
\end{equation}
In our framework the first two terms in the perturbative series of 
the $\beta$ function are scheme-independent. To compute $\beta^{\sMOM}_2$ 
the knowledge of the two loop vertex corrections is sufficient if 
$\beta^{\sMSbar}_2$ is given.

$\beta^{\sMSbar}$ is known up to four loops \cite{beta1l,beta2l,beta3l,beta4l} 
and the first three terms read:
\begin{eqnarray}
\beta^{\sMSbar}_0 &=& \frac{1}{4} 
  \left( \frac{11}{3} C_A - \frac{4}{3} T{} n_f
  \right), \nonumber\\
\beta^{\sMSbar}_1 &=& \frac{1}{16} 
  \left( \frac{34}{3} C_A^2 - 4 C_F T{} n_f - \frac{20}{3} C_A T{} n_f 
  \right), \nonumber\\
\beta^{\sMSbar}_2 &=& \frac{1}{64} 
  \bigg( \frac{2857}{54} C_A^3 + 2 C_F^2 T{} n_f 
       - \frac{205}{9} C_F C_A T{} n_f - \frac{1415}{27} C_A^2 T{} n_f 
\nonumber\\&&\mbox{\hspace{1cm}}
       + \frac{44}{9} C_F T{}^2 n_f^2 + \frac{158}{27} C_A T{}^2 n_f^2
  \bigg).
\end{eqnarray}

The vertex corrections given above lead to the following values for the
three loop $\beta$ function in the $\MOM$-scheme:
\begin{eqnarray}
   \beta_2^{\sMOMg}   &=& 44.82(5) - 9.730(5) \, n_f + 0.3276(1) \, n_f^2, 
\nonumber\\ 
   \beta_2^{\sMOMq}   &=& 28.86(3) - 9.206(3) \, n_f + 0.35322(7) \, n_f^2,
\nonumber\\
   \beta_2^{\sMOMggg} &=& 24(2) + 0.04(63) \, n_f - 1.05(3) \, n_f^2 
                                 + 0.0415330 \, n_f^3.
\end{eqnarray}
Note, that a large  uncertainty of the $n_f$-coefficient for the 
$\MOMggg$-scheme is the result of an accidental cancellation between 
$\beta_2^{\sMSbar}$ and the corrections induced by the one and two loop vertex
corrections in this case.

It is instructive to compare the $\MOMggg$-scheme with the (standard) 
$\MSbar$-scheme as well as with the $\MOMgg$ one. The latter is defined 
with the help of the asymmetric subtraction of the triple gluon vertex (more 
details in \cite{hep-ph/9810437,MOMgg:4loop}). The three-loop contributions 
to the corresponding $\beta$-functions read 
\begin{eqnarray}
   \beta_2^{\sMSbar}   &=& 22.3203 - 4.36892  \, n_f + 0.0940394  \, n_f^2 , 
\nonumber\\
   \beta_2^{\stMOMgg} &=& 
37.6899 - 5.57013  \, n_f - 0.223177  \,  n_f^2  + 
                   0.0138889 \,  n_f^3 
{}.
\end{eqnarray}

Thus, one  observes that (at three loops) the $\MOMggg$-scheme is numerically
significantly closer to the $\MSbar$-scheme that to the $\MOMgg$ one. 

Very recently, in \cite{Boucaud:2000nd} an attempt has been made to "predict"
the value of $\beta_2^{\sMOMggg}|_{n_f=0}$ with the help of an OPE analysis of
the flavourless non-perturbative gluon propagator and the symmetric triple 
gluon vertex in the Landau gauge. They have found 
\begin{equation}
\beta_2^{\sMOMggg}|_{lattice} = 1.5(3) \, \beta_2^{\MOMgg}|_{n_f=0}
\label{they}
\end{equation}
while our result is 
\begin{equation}
\beta_2^{\sMOMggg}|_{n_f=0} = 0.64(5) \, \beta_2^{\MOMgg}|_{n_f=0}
\label{we}
{}.
\end{equation}

We believe that the rough (that is in the sign and overall magnitude) agreement 
between Eqs.~(\ref{they}) and (\ref{we}) could be improved by taking into 
account anomalous dimensions of power corrections as discussed in 
\cite{Boucaud:2000nd}.

\section*{Conclusion} 

We have presented a new approach to approximately compute three-point Green 
functions in massless theories. The approach is based on the use of asymptotic
expansion methods. It reduces the problem of computing of a three-point N-loop
correlator (with Euclidean external momenta) to the calculation of the N-loop
massless propagators. We have checked the method by recomputing a few 
analytically known results. In all cases the resulting series proved to be 
accurate approximations to the exact results. 

We have applied the approach to compute the two loop QCD corrections to the 
ghost gluon vertex, the quark gluon vertex, and the triple gluon vertex in the 
massless limit at the symmetric momentum point. The results have been used to 
construct the corresponding three-loop $\beta$ functions. In principle, the 
calculation could be extended to include one more loop (as the three-loop 
massless propagators are still accessible with {\tt MINCER}). Unfortunately, 
for the time being the hardware constraints look insurmountable.
 
\section*{Acknowledgements} 

We would like to thank J.H.~K\"uhn for numerous fruitful discussions and 
A.I.~Davydychev for providing us with information about the results 
of~\cite{davyd96}. We are grateful to the authors of the work 
\cite{Boucaud:2000nd} for sending us a draft version of it.

This work was supported by DFG under contract FOR 264/2-1, the 
{\it Graduiertenkolleg ``Elementarteilchenphysik an Beschleunigern''} and the
{\it DFG-Forschergruppe ``Quantenfeldtheorie, Computeralgebra und
Monte-Carlo-Simulationen''}.

\def\app#1#2#3{{\it Act.~Phys.~Pol.~}{\bf B #1} (#2) #3}
\def\apa#1#2#3{{\it Act.~Phys.~Austr.~}{\bf#1} (#2) #3}
\def\cmp#1#2#3{{\it Comm.~Math.~Phys.~}{\bf #1} (#2) #3}
\def\cpc#1#2#3{{\it Comp.~Phys.~Commun.~}{\bf #1} (#2) #3}
\def\epjc#1#2#3{{\it Eur.\ Phys.\ J.\ }{\bf C #1} (#2) #3}
\def\fortp#1#2#3{{\it Fortschr.~Phys.~}{\bf#1} (#2) #3}
\def\ijmpc#1#2#3{{\it Int.~J.~Mod.~Phys.~}{\bf C #1} (#2) #3}
\def\ijmpa#1#2#3{{\it Int.~J.~Mod.~Phys.~}{\bf A #1} (#2) #3}
\def\jcp#1#2#3{{\it J.~Comp.~Phys.~}{\bf #1} (#2) #3}
\def\jetp#1#2#3{{\it JETP~Lett.~}{\bf #1} (#2) #3}
\def\mpl#1#2#3{{\it Mod.~Phys.~Lett.~}{\bf A #1} (#2) #3}
\def\nima#1#2#3{{\it Nucl.~Inst.~Meth.~}{\bf A #1} (#2) #3}
\def\npb#1#2#3{{\it Nucl.~Phys.~}{\bf B #1} (#2) #3}
\def\nca#1#2#3{{\it Nuovo~Cim.~}{\bf #1A} (#2) #3}
\def\plb#1#2#3{{\it Phys.~Lett.~}{\bf B #1} (#2) #3}
\def\prc#1#2#3{{\it Phys.~Reports }{\bf #1} (#2) #3}
\def\prd#1#2#3{{\it Phys.~Rev.~}{\bf D #1} (#2) #3}
\def\pR#1#2#3{{\it Phys.~Rev.~}{\bf #1} (#2) #3}
\def\prl#1#2#3{{\it Phys.~Rev.~Lett.~}{\bf #1} (#2) #3}
\def\pr#1#2#3{{\it Phys.~Reports }{\bf #1} (#2) #3}
\def\ptp#1#2#3{{\it Prog.~Theor.~Phys.~}{\bf #1} (#2) #3}
\def\sovnp#1#2#3{{\it Sov.~J.~Nucl.~Phys.~}{\bf #1} (#2) #3}
\def\tmf#1#2#3{{\it Teor.~Mat.~Fiz.~}{\bf #1} (#2) #3}
\def\yadfiz#1#2#3{{\it Yad.~Fiz.~}{\bf #1} (#2) #3}
\def\zpc#1#2#3{{\it Z.~Phys.~}{\bf C #1} (#2) #3}
\def\ppnp#1#2#3{{\it Prog.~Part.~Nucl.~Phys.~}{\bf #1} (#2) #3}
\def\ibid#1#2#3{{ibid.~}{\bf #1} (#2) #3}
\def\jhep#1#2#3{{\it JHEP~}{\bf #1} (#2) #3}

\end{document}